\documentstyle[12pt]{article}

\begin{document}
\centerline {\bf Generalized Hartree Method: A Novel Non-perturbative
 Approximation}
\centerline {\bf Scheme for Interacting Quantum systems }
\medskip
\centerline {B P Mahapatra, $^{\star}$  N Santi $^{\dagger}$ $ {\&}$  N B Pradhan $^{\dagger}$}                    

\centerline {$^{\star}$Department of Physics,Sambalpur University,JyotiVihar 768019}                                                   

\centerline {$^{\dagger}$Department of Physics, G M College(Autonomous), Sambalpur 768004                                                      }
             \smallskip
\begin{abstract}
A  self-consistent ,  non-perturbative  scheme of  approximation is
proposed for arbitrary interacting quantum systems by generalization
of the Hartree method.The scheme consists in 
approximating the original interaction term $\lambda H_I$ by a
suitable 'potential' $\lambda V(\phi)$  which satisfies the following 
two requirements: (i) the 'Hartree  Hamiltonian'  $H_o$ generated by 
$V(\phi)$ is {\sl exactly solvable } i.e, the eigen states  $|n>$  and 
the eigenvalues  $E_n$  are known and (ii) the 'quantum  averages' of
the two are equal, i.e. $ < n|H_I|n> $ = $<n|V(\phi)|n> $  
for arbitrary $'n~'$. The leading-order results  for $|n>$  and  $E_n$ , 
which are already accurate, can be systematically
improved further by the development of a 'Hartree-improved
perturbation theory'(HIPT) with $ H_o$  as the unperturbed part and 
the modified interaction:$\lambda H^{\prime}  \equiv  \lambda (H_I-V)$
as the perturbation. The HIPT is assured  of rapid convergence
because of the 'Hartree  condtion' : $<n| H'| n> = 0 $. 
This is in contrast to the naive perturbation theory developed  with 
the original interaction  term  $\lambda H_I$    chosen as the perturbation,
 which diverges even for infinitesimal $\lambda$ !  The structure of 
the Hartree vacuum is shown to be highly non-trivial.
Application of the method to the anharmonic-and double-well 
quartic-oscillators, anharmonic- sextic and octic- oscillators leads 
to  very accurate results for the energy levels. In case of 
$ \lambda \phi^{4}$ quantum field theory,
the method reproduces, in the leading order, the results of Gaussian 
approximation, which can be improved further by the HIPT. We study the
vacuum structure, renormalisation and stability of the theory in GHA.
\end {abstract}
\newpage
\noindent {\bf 1.   Formulation }

     Consider a generic Hamiltonian describing an arbitrary interacting quantum
system:
\begin{equation}
           H~=~H_s+\lambda H_I(\phi),    
\label{f1}
\end{equation} 
     where  $H_s$  is exactly solvable, $\lambda  H_I(\phi)$  is the interaction  and   $\lambda$  is the strength of interaction. The generalized
Hartree approximation (GHA) to the above Hamiltonian is:
\begin{equation}
                  H_o~ \equiv ~H_s+ \lambda V(\phi)     
\label{f2}
\end{equation} 
 where  $ V(\phi) \equiv$   ' Hartree   potential ' (HP) [ 1 ]  is the approximation  to the original interaction  $H_I (\phi)$  and is required to satisfy the following {\sl two} conditions: (i) $H_o$  ( $\equiv$ ' Hartree  Hamiltonian '(HH))  is exactly solvable,i.e.,
\begin{equation}
          H_o|n>~=~E_n|n>, ~ <m|n> = \delta_{mn},
\label{f3}
\end{equation}      
with the eigen-spectrum known and (ii) the 'quantum average'(QA) of  $V$  equals
that of $H_I,$ [ 1 ]  i.e,
\begin{equation}
<n  |V (\phi)  |n>~=~<n  |H_I (\phi)  |n>
\label{f4}
\end {equation}
(The    quantum   average   of   any   operator   $  <\hat{A} (\phi)>  $   is    henceforth denoted by the    notation   : \\
$ <\hat{ A }>  \equiv  <n| \hat{ A} |n>$ ).
Equations (2)-(4) defining the General Hartree Approximation (GHA)  are hence forward referred to as  "Hartree Conditions (HC)".
The {\sl self-consistency} of the procedure is implicit in eqs.(2-4): the states $|n>$ which are obtained as the solution of $H_0$, are used to specify 
$V (\phi)$ which, in turn, defines $H_0$. The leading approximation consists in finding the spectrum: $|n>$ $and$ $E_n$. {\sl It may be emphasized that even the 
leading- order results capture the dominant contribution of the interaction 
through the requirement, eq.(4),  while one always deals with an exactly solvable
Hamiltonian}.

Note that, because of eq.(4), the QA of $ H^{\prime}$  vanishes:
\begin{equation}
<n|H^{\prime}|n> = 0
\label{f5}
\end{equation}
where $ H'\equiv H_{I}~-~V $.

This naturally suggests a scheme of {\sl improved} perturbation theory
($\equiv$ 'Hartree-improved perturbation theory'  (HIPT)) in which $H_0$  is used as the 
unperturbed part and $ \lambda H^\prime $ is considered as the perturbation (see, below).
\newpage
\noindent{\bf2.  Applications: (Quantum Mechanics) }

\noindent{\bf2.   (a) The quartic - oscillator }

Consider the Hamiltonian for the (quartic) anharmonic oscillator (AHO) and the 
double-well oscillator(DWO) :
\begin{equation}
H~=~\frac{1}{2}p^{2}~+~\frac{1}{2}~g~\phi^{2}+\lambda\phi^{4},
\label{f6}
\end{equation}
where $~g~> ~0~ (< ~0~ )$   refers  to the cases of the AHO(DWO) respectively.  These  systems  have  been widely studied in the literature owing to their theoretical importance as well as
practical applications.
The free-field $\phi(t)$ and  the conjugate momentum p(t) are  parametrised in the standard manner:
\begin{equation}
\phi(t)~=~\sigma+\frac{(b+b^\dagger)}{\sqrt{2\omega}} ,
~ p(t)~=~i \sqrt {\frac {\omega}{2}} (b^\dagger-b),
\label{f7}
\end{equation}
 satisfying  the usual commutaion relation  $[b,b^\dagger] = 1 $.
Here, $ \omega$,  $ \sigma$ are real constants with  $ \omega >0 $ and  $ \sigma $
has the significance of the vacuum  expectation value (VEV) of  $\phi$: $\sigma \equiv <\phi>$.  It is to be noted that free-field operators defined by :
\begin{equation}
\phi~=~\sigma +\frac{1}{\sqrt{2}} (a+a^{\dagger}),
~ p=\frac{i}{\sqrt{2}}(a^{\dagger}-a)
\label{f8}
\end{equation}
satisfy identical  commutaion relation :
$[ a , a^{\dagger}]~=~1.$
Hence $ ( a ,a^{\dagger} )$  and $ (b , b^{\dagger})$ must be  related by Quantum canonical Transformation  i.e.(Boguliobov-like Transformation). Denoting  the eigen states of the number operator  $b^\dagger  b$  as   $|n>$  we have the standard definition
$|n>~=~(b^{\dagger})^{n}|vac>/ \sqrt{ n!}$
 where  $b|vac> \equiv  0$ ; $  b^{\dagger}b|n>  = n|n>$ and $<m|n> = \delta_{mn}$.
It is to be noted  that free field  vacuum $|0>$ is defined as :  $ a|0>\equiv  0 \equiv  <0|a^{\dagger} $ . The QA  are easily  calculated: 
  $ <\phi> =  \sigma $  ;  $ <\phi^{2}> = \sigma^{2} + \xi/\omega $; $ <\phi
^{3}> = \sigma^{3} + 3 \sigma \xi/ \omega $ and  $ <\phi^ {4}> = \sigma^ {4} + 6 \sigma^ {2} \xi / \omega + (3/(8 \omega^{2})) ( 1 + 4 \xi^{2})$ ; where $ \xi \equiv n + 1/2 $.

To set up the  HH for the above system we propose the generic {\sl ansatz}
({\sl valid for arbitrary  anharmonicity}):
\begin{equation}
V(\phi)~=~A \phi^{2} - B \phi + C
\label{f9}
\end{equation}
which is amenable to immediate implementation of the HC. {\sl Self-consistency}
of the procedure is built in by imposing the constraint that the constants
A,B,C are chosen as suitable functions of  $<\phi^{n}>$ . For  the case of {\sl quartic} anharmonicity, we determine  them as follows:
\begin{eqnarray}
&A&  =  6  \sigma^ {2} + 3 f( \xi )/ \omega ; ~B~=~(1+g) ~\sigma\omega^{2}/\lambda~~+~4~\omega^{2}~\sigma^{3}~+12~\omega~\sigma~\xi ;
\nonumber \\
&C&=~<\phi^{4}>~~-~A~<\phi^{2}>~+~B~<~\phi~>
\label{f10}
\end{eqnarray}
 where  $ f ( \xi ) =  \xi + (1/ 4 \xi )$ . Note that, we have imposed the  {\sl additional} constraint of reproducing the variational " Gap- equation" and the "equation for the ground state" in the Gaussian-approximation  [  2  ], in determining the constants: A,B,C.   
With the choice above, the HH is reduced to the diagonalisable-structure,
corresponding to a shifted - harmonic oscillator:
\begin{equation}
H_0~=~h_0 + \frac{1}{2} \Omega^{2}(\phi-\chi)^{2} + \frac{1}{2} p^{2},
\label{f11}
\end{equation}
where $ \Omega^{2}~=~2 \lambda A +~g,~ \chi~=~\lambda B/ \Omega^{2}$
and   $  h_0~=~\lambda C - \frac{1}{2}\Omega^{2} \chi^{2}.$
However, demanding the consistency of the diagonalisation of $H_0$ to the defining {\sl ansatz} eq.(8), requires that we make the identification :
\begin{equation}
\omega^{2}~=~\Omega^{2} ; ~\sigma^{2}~=~\chi^{2},
\label{f12}
\end{equation}
which, in turn, imply the following constraints on  $~\omega~$  and  $~\sigma~$:
\begin{equation}
 \omega^{3} - \omega ~( 12 \lambda \sigma^ {2}
 + g ) - 6 \lambda f(\xi) = 0
\label{f13}
\end{equation}
\begin{equation}
\sigma ~(  4 \lambda \sigma^ {2} + g + (12 \lambda \xi / \omega) ) = 0
\label{f14}
\end{equation}
(Here inafter,eqs.(13 ) $ \&$  (14 ) are referred to as the  "gap-eqn (GE)" and the "equation for ground state" ( EGS ) respectively). (As stated earlier, $g > 0 ~( < 0 )$  refers to the case of the AHO(DWO)). In the case of AHO the  $\sigma~=~0,$  solution  of
eq(14 ) corresponds  to the {\sl physical} solution. This leads to the simplified GE, ( g = 1):
\begin{equation}
\omega^ {3} - \omega - 6 \lambda f (\xi) = 0.
\label{f15}
\end{equation}
 This gap-equation has been  derived  earlier  by several authors [3 - 6 ]  from a variety of different considerations. With the aid  of eqs.(11) $\&$  (15), the 
energy-levels of $H_0$ are given by
\begin{equation}
 E_{n} ~=~\frac{\xi}{4} (3\omega+\frac{1}{\omega}),
\label{f16}
\end{equation}
where $ '\omega' $ is the  solution of eq.(15).
 The  numerical  results  for  $ E_{n}$ 
 are  given in $\bf Table-1 $.
The result is already accurate with error   $ ~ \sim  0.2 ~\% ~ to  ~2 ~\% ~$ over the range of $'n'$ \&  $'\lambda'$ shown.

 For the case of the DWO,  there are two  ${~\sl quantum ~phases}$:
( i ) $ 4 \lambda \sigma^ {2} = - g  - (12 \lambda \xi / \omega) $, that leads to the
spontaneously broken symmetry (SSB) phase and  
 (ii) $\sigma = 0 $, that leads to Symmetry-Restored phase ( SR - phase ) (see,eqn.( 14 ).
 The former exhibits the double-well structure of the 'effective-potential'(for which case, the VEV of the field is non-vanishing: $\sigma^{2}\neq 0$) whereas the latter 
 corresponds to the dynamical  restoration of the single-well shape, with $\sigma~=~0$. 

The gap equations in respective phases are easily obtained :
for the {\sl SSB phase} ( see eq.(13) )
\begin{equation}
\omega^ {3}_a ~+~ 2 g~ \omega_a + 6 \lambda p(\xi) = 0
\label{f17}
\end{equation}
where $ p (\xi)\equiv (5 \xi - 1/4 \xi)$
with the physical solution given by :
\begin{equation}
  \omega_{a}~=~2\sqrt (\frac{-2g}{3})
cos ~[ \frac{\pi}{6}+(\frac{1}{3})sin^{-1}(\frac{\lambda}{\lambda_c})
~ ]
\label{f18}
\end{equation}
\begin{equation}
 \lambda_{c}(\xi) = ( -2g/3)^{3/2} / 3 p(\xi)
\label{f20}
\end{equation}
where $\lambda_c$ = "critical coupling".
 For the  ground state ( n = 0, g= - 1,   $\xi = 1/2$) :
$ \lambda_c (1/2)\simeq 0.09007 $. This shows that the SSB-phase is realized
for $ \lambda < \lambda_c $ and the  SR phase is favoured  for $\lambda > \lambda_c$.
The energy levels in SSB -phase are given by
\begin{equation}
 E_{n}^{SSB} = ~\frac{\xi}{4} (3~\omega_{a} + \frac{2}{\omega_{a}}~) - \frac{g^{2}}{16\lambda} 
\label{f21}
\end{equation}
For the {\sl SR phase}: 
\begin{equation}
\omega^ {3}_s -~g~ \omega_s - 6 \lambda f (\xi) = 0
\label{f22}
\end{equation}
The corresponding expression for the energy-levels, are given by :
\begin{equation}
 E_{n}^{SR} = (\xi / 4) [ 3 \omega_s + g / \omega_s ]
\label{f23}
\end{equation}
where ${\omega_{s}}$ is the solution of eqn.(21) and ${\lambda}~ >$ ${\lambda_{c}}$
 (see,{\bf Table-2}).
\newpage
\noindent {\bf2. (b) Cases of higher harmonicity}

Other cases of anharmonicity  are studied in analogous manner. For the  anharmonic sextic- oscillator, the Hamiltonian  is given by
\begin{equation}
H~=~\frac{1}{2}p^{2}~+~\frac{1}{2}~g~\phi^{2}+\lambda\phi^{6}
\label{f24}
\end{equation}
where $g~>~0~(<~0)$ refers to the case of AHO (DWO) respectively.
 The  Hartree-Hamiltonian for this case is chosen as:
\begin{equation}
H_{0}~=~\frac{1}{2}p^{2}~+~\frac{1}{2}~g~\phi^{2}+\lambda~V(\phi)
\label{f25}
\end{equation}
with the condition,
\begin{equation}
<H>~=~<H_{0}>
\label{f26}
\end{equation}
Identical ansatz, eq.(9), for  $V(\phi)$ is assumed:
\ $ V(\phi)~=~A~\phi^{2}~-~B\phi~+~C$. The constants are calculated in the analogous manner and    given as:
\begin{eqnarray}
&A&=~15~\sigma^{4} +45\sigma^{2}(4\xi^{2}+1)/4\xi\omega +(15/8\omega^{2})
(4\xi^{2}+5); 
\nonumber\\
&B&=~\sigma~[(1+g)\omega^{2}/\lambda
+6\omega^{2}\sigma^{4} +60\sigma^{2}\xi\omega +(45/4)(4\xi^{2}+1)~];
\nonumber\\
&C&=~<\phi^{6}>-A~<\phi^{2}> +B~<\phi>
\label{f27}
\end{eqnarray}
With the above choice the  Hatree-Hamiltonian, $H_{0}$ is again  reduced to  the diagonalisable form  with a shiftted field $\tilde\phi$= $\phi-\sigma$.
The  structure of HH is identical to eq.(11); together with eq.(12): 
\begin{equation}
H_{0}~=~\frac {p^{2}}{2}~+~\frac {1}{2}\omega^{2}(\phi-\sigma)^{2}+h_{0}
\label{f28}
\end{equation}
where $ \omega^{2}~=~2 \lambda A +~g,$~$ \sigma~=~\lambda B/ \omega^{2}$
and   $  h_0~=~\lambda C - \frac{1}{2}\omega^{2} \sigma^{2}.$
The 'gap -eqn' in this case is given by
\begin{equation}
\omega^{4}~-\omega^{2}(g+30\lambda\sigma^{4})-45\lambda(\sigma^{2}\omega/2\xi)(4\xi^{2}+1) -(15\lambda/4)(4\xi^{2}+5) = 0
\label{f29}
\end{equation}
whereas, the ground-state-configuration is governed by the following equation: 
\begin{equation}
\sigma~[ g+6\lambda(\sigma^{4}+10\xi\sigma^{2}/\omega+15(4\xi^{2}+1)/8\omega^{2})~]=0.
\label{f30}
\end{equation}

In the case of AHO, the  $\sigma~=~0 ~$  solution of eqn. (29) corresponds to the $\sl physical$ solution. This leads to the simplified "gap-eqn.":
\begin{equation}
\omega^{4}-g\omega^{2} -(15\lambda/4)(4\xi^{2}+5) = 0
\label{f31}
\end{equation}
By using eqs.(27) and(30) the energy levels of $H_{0}$ are given by
\begin{equation}
 E_{n}=\frac{\xi}{3}(2\omega+\frac{g}{\omega})
\label{f32}
\end{equation}
where '$\omega$' is the solution of eq(30).The numerical results for  $  E_{n}$ are given in {\bf Table-3}.

For the case of the  octic-anharmonic oscillator the Hamiltonian and the HH of the system are  given respectively  by
\begin{equation}
H~=~\frac{1}{2} p^{2}+\frac{1}{2}g\phi^{2}+\lambda\phi^{8},
\label{f33}
\end{equation}
\begin{equation}
and~~~~~~~      H_{0}~=~\frac{1}{2} p^{2}+\frac{1}{2}g\phi^{2}+\lambda V(\phi),
\label{f34}
\end{equation}
with  identical ansatz for $V(\phi)$:

       $V(\phi)~=~A~\phi^{2}~-~B~\phi~+~C$.
The constants A,B,C are analogously determined:
\begin{eqnarray}
&A&=~28\sigma^{6} +105\sigma^{4}(4\xi^{2}+1)/2\xi\omega +(105/2\omega^{2})\sigma^{2}
(4\xi^{2}+5)+35 h(\xi)/2\omega^{3} ;
\nonumber\\
&B&=~\sigma~[(1+g)\omega^{2}/\lambda
+8\omega^{2}\sigma^{6} +168\sigma^{4}\xi\omega +105\sigma^{2}(4\xi^{2}+1)+35\xi(4\xi^{2}+5)/\omega~];
\nonumber\\
&C&=~<\phi^{8}>~-~A~<\phi^{2}> ~+~B~<\phi>
\label{f35}
\end{eqnarray}

where $ h(\xi)~=~\xi^{3}+(7\xi/2)+(9/16\xi)$.

 The " gap-eqn" and eqn. for "ground- state" are as follows: 
\begin{equation}
\omega^{5}-\omega^{3}(g+56\lambda\sigma^{6})-105\omega^{2}(\lambda\sigma^{4}/\xi)(4\xi^{2}+1)-105\omega\lambda\sigma^{2}(4\xi^{2}+5)-35\lambda h(\xi)=0,
\label{f36}
\end{equation}
\begin{equation}
\sigma~[ g+\lambda(8\sigma^{6}+168\sigma^{4}(\xi/\omega)+105\sigma^{2}(4\xi^{2}+1)/\omega^{2}+35\xi(4\xi^{2}+5)/\omega^{3})~]
\label{f37}
\end{equation}
For the case of anharmonic-octic oscillator  the solution   $\sigma ~=~0$ is the  physical solution. In  this case the "gap-eqn" is simplified to
\begin{equation}
\omega^{5}-g\omega^{3}-35\lambda h(\xi)=0
\label{f38}
\end{equation}
which leads to the energy-levels of $H_{0}$ 
\begin{equation}
 E_{n}=(\frac{\xi}{8})(5\omega+\frac{3g}{\omega})
\label{f39}
\end{equation}
with '$\omega$'as the solution of eqn(37).The numerical results are given in {~\bf Table-4~}, compared with earlier results [ 7 ].

The generality  of the above method is thus apparent  from the above examples of increasing anharmonicity. In each case , the interacting system has been $\sl effectively$ reduced to an exactly solvable system while preserving the inherent non-linearity (through the gap-eqn.$\&$ the Hartree-condtion) of the interacting theory. The resulting energy- levels are $\sl uniformly$ accurate to within a few percent of the 'exactly' computed values $~\sl even~$  in the zeroth-order, which is chosen to  reproduce the results of the gaussian-approximation [ 2 ]. Further physical significance of the GHA is obtained by studying the structure and stability of the Hartree-vacuum as discussed  below.

\noindent  {\bf3.   Structure and significance of the Hartree-vacuum}

Starting from the alternative expansion of the field in terms of the 'free'-field
operators,
\begin{equation}
\phi~=~\sigma + ( a + a^{\dagger})/ \sqrt{2} ;
~~p~=~( i/ \sqrt{2} )(a^{\dagger} - a ),
~[ a,a^{\dagger}]~ =~1,
\label{f40}
\end{equation}
it follows (from eqs.(7))  that the two sets of operators must be related by a Boguliobov-type quantum canonical transformation :
\begin{eqnarray}
&b&~=~a~ \cosh(\alpha) - a^{\dagger}\sinh (\alpha)
\nonumber \\
&b^{\dagger}&~=~a^{\dagger}\cosh(\alpha) - a~  \sinh (\alpha),
\label{f}
\end{eqnarray}
with the two vacua related by the transformation
\begin{eqnarray}
&|vac>& = exp [(1/2)~ \tanh (\alpha)~ (a^{\dagger} a^{\dagger} - a a)] |0>~ ;
\nonumber \\
&\alpha&\equiv  (1/2) ~ ln (1/ \omega)
\label{f41}
\end{eqnarray}
The set of equations: (40 ,41) imply a highly non-trivial structure of the 
Hartree-vacuum analogous to the case of the super-fluid ground state 
[ 8 ]  and  the   hard-sphere-Bose gas $~ [~ 9 ~].~$ In particular, the 
free-particle number-density in the Hartree-vacuum is non-zero and depends 
strongly on the strength of interaction:
\begin{equation}
n_0 \equiv < vac|a^{\dagger} a| vac > = \sinh^{2} (\alpha) = (1/4) 
[\omega + (1/ \omega) - 2 ]
\label{f42}
\end{equation}
such that  $ n_0\sim\lambda^{1/3} $ for $ \lambda~>>~1$  whereas $n_0\rightarrow 0$ for $\lambda\rightarrow 0 $ (as expected). The above result also implies an alternative interpretation of $'\omega'$, i.e;  $ u\equiv ( 1-\omega)~/( 1+ \omega )$     measures the non-trivial structure [ 10 ]  of the Hartree-vacuum. Moreover, it can be shown [ 10 ] that the {\sl free-field ground state, $|0>$  is highly unstable compared to the Hartree-vacuum, $|vac>$ signifying that the true vacuum is much better approximated by the latter}.

\noindent  {\bf4.   Perturbative Improvement}

An improved perturbation theory (HIPT) can be developed by treating the Hartree
Hamiltonian $H_0$ as the unperturbed part and $\lambda H^\prime$ as the perturbation. The HIPT is, by construction (because of eq.(4)), guaranteed to be convergent, in contrast to the ordinary perturbation theory (wherein the entire interaction  $\lambda H_I$  is treated as the perturbation) which is divergent [ 11-13 ] even for infinitesimal $\lambda$. The first- order correction in HIPT  vanishes and the $2^{nd}$ order -correction  is given by the standard formula : 
\begin{equation}
\Delta E^{(2)}_{n}~=~\sum_{m\neq n} |<n |\lambda H^{\prime}|n> |^{2}
/( E_{n} - E_{m})
\label{f43}
\end{equation}
Only a  finite  number of  matrix-elements  contribute to eq.(43 )  corresponding to 
m = $~n \pm 2, n \pm 4~$ for the case of the quartic AHO (DWO) . Inclusion of this correction significantly improves
the zeroth-order results, both in case of the AHO ({\bf Table-1}) and the DWO ({\bf Table-2}), over the full range of investigation: $0.1 \leq \lambda \leq 100$ and
$ 0 \leq n\leq 40$.

\noindent   {\bf5.   $\bf \lambda \phi^{4}$ - Quantum Field Theory}

The above formalism is easily extended to $\lambda \phi^{4}$ quantum field theory described by the Lagrangian: $ {\cal  L}~=~\frac{1}{2}(\partial_\mu\phi)
(\partial^{\mu}\phi) - \frac{1}{2}m^{2}\phi^{2} - \lambda\phi^{4}$
which leads to the Hamiltonian:
\begin{equation}
{\cal H}~=~\frac{1}{2}~[(\partial\phi/\partial t)^{2} + (\nabla\phi)^{2}
+ m^{2}\phi^{2})~]+ \lambda\phi^{4}
\label{f44}
\end{equation}
The field $\phi$ is Fourier-expanded in terms of the 'interacting' Fock-space operators in the standard manner:
\begin{equation}
\phi~=~\sigma + \int\frac{d^{3}\bf k}{\Omega_{k}(M)}~[b({\bf k})exp(-ikx)
+b^{\dagger}({\bf k})exp(ikx)~]
\label{f45}
\end{equation}
where, $\Omega_{k}(M) \equiv (2\pi)^{3} 2\sqrt{k^{2}+M^{2}} = (2\pi)^{3}2\omega_{k}
(M)$,  $~ [ b(k),b^\dagger(q)~]  ~=~ \Omega_k(M)\delta^{3}(\bf k-\bf q)$,   $~~M~ =~$ bare mass of the ${\sl physical}$ particle and $|vac>$ is defined by  $b({\bf k})|vac> = 0$.  Similarly, the ${\sl physical}$ one-particle state is 
defined by  $|{\bf k}> \equiv b^{\dagger}({\bf k})|vac>.$  The Hartree-Hamiltonian is constructed  in an analogous manner:
\begin{equation}
{\cal H}_0~=~\frac{1}{2}~[(\partial\phi/\partial t)^{2} + (\nabla\phi)^{2}
+ m^{2}\phi^{2})~]+ \lambda V(\phi)
\label{f46}
\end{equation}
with   $V(\phi)$ parameterised as before :
\begin{equation}
V(\phi)~=~A ~\phi^{2} ~- ~B \phi ~+~C 
\label{f47}
\end{equation}

The various  QA's are now given by : $<\phi^{2}>=\sigma^{2}+I_0,$
$<\phi^{4}>=\sigma^{4}+6\sigma^{2}I_0+3I_0^{2},$
$<(\partial\phi/\partial t)^{2}>=I_1,$ $<(\nabla\phi)^{2}>=I_1-M^{2}I_0$,
            where $<\hat A>$$\equiv$  $<vac|\hat A|vac>$ and
\begin{equation}
I_n~=~I_n(M^{2})\equiv\int\frac{d^{3}k}{\Omega_k(M)}~[\omega_{k}^{2}
(M)~]^{n},~  n = 0,\pm1,\pm2,\pm3,...,
\label{f48}
\end{equation}
are the Stevenson-Integrals  [ 14 ]. Applying the Hartree-condition
, eq.(4), and ensuring equivalence to the GEP [ 14 ]  leads, as before, to the complete determination of the parameters in $V(\phi)$: 
\begin{equation}
A~=~6~<\phi^{2}>~;~B~=~8~\sigma^{3} ~~and~~ ~C~=~3\sigma^{4}~-~6\sigma^{2}I_{0}~-~3I_{0}^{2}
\label{f49}
\end{equation}
By  analogous   procedure, the  Hartree - Hamiltonian is then  reduced to diagonal form  (corresponding to an $\sl~ effectively~ free~$    Klein-Gordon theory) by a shift of the field and energy :
\begin{equation}
{\cal H}_0~=~\frac{1}{2}~[(\partial\varphi/\partial t)^{2} + (\nabla\varphi)^{2}
+ M^{2}\varphi^{2})~]+h_0
\label{f50}
\end{equation}
where  $\varphi\equiv\phi-\sigma$ and  $h_0~=~-(1/2) M^{2}\sigma^{2}~+~\lambda~C,$ with 'C' given in eq.( 49 )
      
The 'gap-equation' is given by 
\begin{equation}
M^{2}~=~m^{2}+12\lambda\sigma^{2}+12\lambda I_0(M^{2})
\label{f51}
\end{equation}
and the VEV now satisfies the $\sl consistency$ condition:
\begin{equation}
\sigma [ M^{2}~-~8~\lambda~\sigma{^2} ]~=~0
\label{f52}
\end{equation}

   The $\sl physical$
solution of this equation for the vacuum-configuration is at $\sigma = 0$, as intuitively expected for the symmetric $\lambda\phi^{4}$ theory (i.e. with $m^{2}~>~0)$  considered here. This is further verified by the computation of the $\sl effective~ potential$ and the $\sl renormalized~  parameters$ as shown below.

Employing the  standard  definition of the $ \sl  effective~ potential $ $U(\sigma)$: 
$<vac|{\cal H}|vac>$ = $<vac|{\cal H}_0|vac> \equiv U(\sigma)$, we have,
\begin{equation}
U(\sigma)=I_1-3\lambda I_0^{2}+(1/2)m^{2}\sigma^{2}+\lambda\sigma^{4}
\label{f53}
\end{equation}
It is important to note that,  eqs.( 51 - 53 )  coincide with the results based on the ${\sl  Gaussian~ Effective~ Potential~ (GEP)~ }$[ 14 ]  . It is then straight- forward to carry out the renormalisation 
program as has been done in ref.[14 ]. The ${\sl renormalized~  parameters} $ are given by the following expressions:
\begin{equation}
m_{R}^{2} \equiv d^{2}U/d \sigma^{2}|_{\sigma=0} = m^{2}+12 \lambda I_0
(\bar {M^{2}}); \bar {M^{2}}\equiv M^{2} (\sigma=0)
\label{f54}
\end{equation}
\begin{equation}
\lambda_{R}  \equiv (1/4!) d^{4}U/d \sigma^{4}|_{\sigma=0} = \lambda \left[   
\frac{1-12\lambda I_{-1}}{1+6\lambda I_{-1}}\right].
\label{f55}
\end{equation}
It has been  shown [14] in case of the GEP that a $ \sl non-trivial~ version $ of $\lambda\phi^{4}$ theory emerges which can $\sl not$ be realized in lattice theories (for any $\sl finite $ lattice-spacing). Identical conclusion holds for the GHA in the leading 
order, because of the equivalence established here.

Beyond  proving the equivalence of the GHA to GEP in the leading order, the new 
 results derived  are :  

(i) the $\sl vacuum-structure$:
\begin{equation}
|vac>=exp\{\frac{1}{2}\int\frac{d^{3}{\bf k}}{\Omega_k(m)}
\beta({\bf k})~[a^{\dagger}({\bf k})a^{\dagger}(-{\bf k})
-a({\bf k})a(-{\bf k})~]\}
|0>
\label{f56}
\end{equation}
where $\beta({\bf k})$  is analogous  to $\alpha({\bf k})$ (see,eqn.(40) with $\alpha~
\rightarrow ~\alpha({\bf k}))$.
The "vacuum- structure-function"  is denoted by $u(\bf k)$ given by 
\begin{equation}
u({\bf k})~=~\omega_k(m)/ \omega_k(M)~=~[({\bf k}^{2}+m^{2})/
({\bf k}^{2}+M^{2})~]^{1/2}
\label{f57}
\end{equation}
The  free-particle  number - density in the Hartree - vacuum is given by 
(V~=~ spatial~\\ volume )
\begin{equation}
n({\bf k}) \equiv <vac|a^{\dagger}({\bf k})a({\bf k})|vac>
/ V \Omega_k(m)
\label{f58}
\end{equation}
One then calculates $ \rho({\bf k})$  as defined  below
\begin{equation}
\lim_ {\Lambda\rightarrow\infty} ~ n({\bf k})/n({\bf 0})\equiv \rho({\bf k})=
~[ 1+({\bf k}^{2}/m^{2}_R)~]^{-1/2}  
\label{f59}
\end{equation}
where $ n({\bf 0})=(m/m_R)/(32 \pi^{3})$ is the maximum value of $n({\bf k})$.
Eqs.( 57-58) show the condensate- structure of the physical vacuum consisting of
correlated off- shell particle pairs ( with momenta $\bf k$ and -$\bf k)$ with a non-trivial dependence on $|{\bf k}|$! This condensate - structure is expected to play significant role in the thermodynamic- properties of the system, if the behaviour persists  to non-zero value of temprature, T.

 The static- potential $~ U({\bf r})~$
 of the ${\lambda\phi^{4}}$ theory is easily calculable from the 2-particle 
correlation function $ U({\bf x - \bf y})$  defined as: 
\begin{equation}
 U({\bf x - \bf y})\equiv <vac|\phi({\bf x},0)\phi({\bf y},0)|vac>/<vac|vac>
\label{f60}
\end{equation}
which leads to
\begin{equation}
U(r)~=~\frac{1}{2}\int\frac{d^{3}{\bf k}}{(2\pi)^{3}}exp(i{\bf k}.{\bf r})
({\bf k}^{2}+m^{2}_{R})^{-1/2}
\label{f61}
\end{equation}
where ${\bf r}\equiv {\bf x}-{\bf y}.$ The above integral can be related to the
modified Bessel-function $K_{1}(m_{R} r)$ and one obtains:
\begin{equation}
U(r )~=~m_{R} K_1(m_{R}  r)/ 4 \pi^{2} r,
\label{f62}
\end{equation}
where, $ r \equiv |{\bf r}|$. Note that $ U(r) $ diverges in the limit $~ r\rightarrow 0, ~$ as it should (in any local quantum field-theory) and behaves asymptotically as 
\begin{equation}
\lim_{r\rightarrow\infty} U(r)\sim r^{-3/2}exp(-m_R r)
\label{f63}
\end{equation}
which shows a $ \sl  fall- off~ faster~ than~ the~ Yukawa~- potential! $  To our knowledge, the important information (contained in Eqs.( 60 - 63 )) regarding the static-potential of the symmetric $\lambda\phi^{4}$ theory are new results.

\noindent {\bf 6.   Summary and Conclusion}

The generalised Hartree approximation is, in principle, applicable to arbitrary quantum-systems with interaction. The method consists in $ \sl mapping~ the~$   $\sl  interacting~ system~ to~ an$
~$\sl  exactly~ solvable~ system $  while preserving the essential non-linearity and the major effects of the interaction of the original system through the self-consistent feed-back mechanism. The basic approximation, which is $ \sl non-perturbative $ in character, lends itself to be systematically improved by modified perturbation theory developed about the Hartree-vacuum and the Hartree-
Hamiltonian chosen as the unperturbed part and  the modified interaction term
$ \lambda H^\prime $   taken as the perturbation. This modified perturbation theory can be shown  to be rapidly  convergent in contrast to the divergence of the naive perturbation theory developed about the free-field vacuum with the      entire 
interaction term $\lambda H_I$ chosen as perturbation. $\sl  We~ conjecture$~ 
~$\sl that~ the~ instability~ of~ the~ free-field~ vacuum~ and~ the~ divergence~~ of$ $\sl ~ perturbation$~
~$\sl theory~ developed~ about~ this~ vacuum~$ $\sl  may~ be~ intimately~ related$ .  The structure of the Hartree-vacuum emerges to be highly non-trivial with a strong dependence on 
$\lambda$ analogous to the case of the ground states of superfluid Helium, hard-
sphere Bose-gas etc.

The method and its perturbative improvement  applied to the case of the 
quartic, sextic and the octic anharmonic - oscillator and the quartic double-well  oscillator  leads to excellent results  over the entire allowed range of the interaction strength $\lambda$ and 
excitation energy level $^\prime n^\prime$. The generalisation to the case of 
$\lambda \phi^{4}$ field theory reproduces, in the leading order,  the results derived from the $ \sl Gaussian~ effective~ potential~$ [~14~]. Going beyond 
the leading order through the HIPT, the results of the Gaussian approximation can be systematically improved. The structure of the Hartree-vacuum again emerges
quite non-trivial and interesting, being characterized by the $ \sl condensation$
of  off-shell, correlated particle-pairs and leading to the definition of a 
$\sl' vacuum-structure~ function'.$ It has  also been established elsewhere [ 10  ]  that the effective potential based upon the free-field ('perturbative') vacuum leads either to instability or to $ \sl triviality $. We also derive the $\sl inter-particle~ (static)~ potential$ by evaluating the two-particle correlation-function in the Hartree vacuum and show  that the potential falls faster than the Yukawa 
potential! The generalization to finite temperature  appears straight-forward.

$\underline{ ACKNOWLEDGEMENT}~~~$  The work is partially supported by a research grant,(No. F.10-83/90(RBB-II)) of University Grants Commission  to BPM.
\newpage
 NS and NBP acknowledge support through UGC- minor research projects (No. F-PS-O-19/97/ERO and no. F-PS-O-18/97/ERO).

\newpage
\begin{table}
\vspace {.333in}
 {\bf Table- 1: }
Sample results  for the zeroth - order (GHA) compared with results of earlier calculations from ref.[ 3 ] (shown in parentheses), over a wide range of $'\lambda'$
and $' n '$. Also shown are the results (in square brackets) of the Hartree -improved perturbation theory (HIPT) up to second order (see text).
\vspace{0.10in}
\begin{center}
\begin{tabular}{c c c c c c c }

\multicolumn{1}{ c }{$\lambda $}&
\multicolumn{1}{ c }{$E_0$ }&
\multicolumn{1}{ c }{\quad$E_1$}&
\multicolumn{1}{ c }{$E_2$}&
\multicolumn{1}{ c }{$E_4$}&
\multicolumn{1}{ c }{$E_{10}$}&
\multicolumn{1}{ c} {$E_{40}$}\\
 
0.1 &0.56031&\quad 1.7734&\quad 3.1382&\quad 6.2052&\quad 17.2267&\quad 94.84
 \\
  &(0.55915)&\quad(1.7695)&\quad(3.1386)&\quad(6.2203)&\quad(17.352)&\quad(90.56)\\
  &[0.55911]&\quad[1.7694]&\quad[3.1391]&\quad[6.2239]&\quad[17.374]&\quad[95.766]\\

1.0&0.81250&\quad2.7599&\quad5.1724&\quad10.902&\quad32.663&\quad192.79\\
  &(0.80377)&\quad(2.7379)&\quad(5.1792)&\quad(10.902)&\quad(32.963)&\quad(194.60)\\
  &[0.80321]&\quad[2.7367]&\quad[5.1824]&\quad[10.982]&\quad[33.013]&\quad[195.15]\\

10.0&1.5313&\quad5.3821&\quad10.3240&\quad22.248&\quad68.177&\quad409.89\\
  &(1.5050)&\quad(5.3216)&\quad(10.3471)&\quad(22.409)&\quad(68.804)&\quad(413.94)\\
 &[1.5030]&\quad[5.3177]&\quad[10.356]&\quad[22.457]&\quad[68.996]&\quad[415.18]\\

100.0&3.1924&\quad11.325&\quad21.853&\quad47.349&\quad145.843&\quad880.55\\
  &(3.1314)&\quad(11.187)&\quad(21.907)&\quad(47.707)&\quad(147.231)&\quad(889.32)\\
  &[3.1266]&\quad[11.178]&\quad[21.927]&\quad[47.817]&\quad[147.652]&\quad[892.03]\\

1000.0&6.8280&\quad24.272&\quad46.902&\quad101.742&\quad313.720&\quad1895.90\\
  &(6.6942)&\quad(23.972)&\quad(47.017)&\quad(102.516)&\quad(${-~~-}$)&\quad(${-~~-}$)\\
  &[6.6836]&\quad[23.952]&\quad[47.062]&\quad[102.75]&\quad[317.65]&\quad[1920.70]\\

\end{tabular}
\end{center}
\end{table}
\begin{table}
\vspace {.4in}
 {\bf Table- 2: }
  The  computed energy levels of the quartic - DWO in the zeroth order of GHA 
compared with the earlier calculations  including twenty - orders of perturbation
 theory $~[ 3 ]~$ shown for sample values of$~'\lambda'~$ and $~'n'~$. Also shown are the
 results after inclusion of  second order correction in HIPT, denoted as
 $E^{(2)}_{n}$.
\vspace{0.13in}
\begin{center}
\begin{tabular}{ c c c c c }

\multicolumn{1}{ c }{$\lambda $}&
\multicolumn{1}{ c }{\quad\quad$ n$ }&
\multicolumn{1}{ c }{\quad\quad$E^{(0)}_{n}$}&
\multicolumn{1}{ c }{\quad\quad$E^{(2)}_{n}$}&
\multicolumn{1}{ c }{ \quad\quad$ Ref[ 3 ]$}\\

0.1 &\quad\quad0 &\quad\quad\quad0.5496&\quad\quad\quad 0.4606&\quad\quad\quad0.4702\\
    &\quad\quad1&\quad\quad\quad0.8430&\quad\quad\quad0.7553&\quad\quad\quad0.7703\\
    &\quad\quad2&\quad\quad\quad1.5636&\quad\quad\quad1.6547&\quad\quad\quad1.6300\\
   &\quad\quad4&\quad\quad\quad3.5805&\quad\quad\quad3.7232&\quad\quad\quad3.6802\\
   &\quad\quad\quad10&\quad\quad\quad12.192&\quad\quad\quad12.517&\quad\quad\quad12.400\\
1.0&\quad\quad\quad0&\quad\quad\quad0.5989&\quad\quad\quad0.5752&\quad\quad\quad0.5800\\
   &\quad\quad\quad1&\quad\quad\quad2.1250&\quad\quad\quad2.0800&\quad\quad\quad2.1800\\
    &\quad\quad\quad2&\quad\quad\quad4.2324&\quad\quad\quad4.2600&\quad\quad\quad4.2500\\
    &\quad\quad\quad4&\quad\quad\quad9.4680&\quad\quad\quad9.5950&\quad\quad\quad9.5600\\
   &\quad\quad\quad10&\quad\quad\quad30.530&\quad\quad\quad30.650&\quad\quad\quad30.420\\
10.0 &\quad\quad\quad0&\quad\quad\quad1.4098&\quad\quad\quad1.3752&\quad\quad\quad1.3800\\
   &\quad\quad\quad1&\quad\quad\quad5.0650&\quad\quad\quad4.9910&\quad\quad\quad5.0900\\
    &\quad\quad\quad2&\quad\quad\quad9.8660&\quad\quad\quad9.9050&\quad\quad\quad9.8900\\
   &\quad\quad\quad4&\quad\quad\quad21.561&\quad\quad\quad21.791&\quad\quad\quad21.700\\
   &\quad\quad\quad10&\quad\quad\quad66.950&\quad\quad\quad67.820&\quad\quad\quad67.620\\
100.0   &\quad\quad\quad0&\quad\quad\quad3.1340&\quad\quad\quad3.0650&\quad\quad\quad3.0700\\
   &\quad\quad\quad1&\quad\quad\quad11.175&\quad\quad\quad11.024&\quad\quad\quad11.002\\
   &\quad\quad\quad2&\quad\quad\quad21.638&\quad\quad\quad21.715&\quad\quad\quad21.700\\
    &\quad\quad\quad4&\quad\quad\quad47.023&\quad\quad\quad47.505&\quad\quad\quad47.200\\
   &\quad\quad\quad10&\quad\quad\quad145.27&\quad\quad\quad147.10&\quad\quad\quad146.70\\
\end{tabular}
\end{center}
\end{table}
\newpage
\begin{table}
\vspace {.4in}
 {\bf Table- 3: }
Sample results  for the zeroth - order (GHA) for the sextic - AHO  compared with results of earlier calculations from ref.[ 7 ] (shown in parentheses), over a wide range of $'\lambda'$
and $'n'$. Percentage of error is shown in square bracket. 
\vspace{0.13in}
\begin{center}
\begin{tabular}{c c c c c c c }
\multicolumn{1}{ c }{$\beta~=  $}&
\multicolumn{1}{ c }{\quad$0.2$ }&
\multicolumn{1}{ c }{\quad$2.0$}&
\multicolumn{1}{ c }{\quad$10.0$}&
\multicolumn{1}{ c }{\quad$100.0$}&
\multicolumn{1}{ c }{\quad$400.0$}&
\multicolumn{1}{ c }{\quad$2000.0$}\\

${E_{0}}$ &1.193&1.676&2.323&3.947&5.521&8.206\\ 
  &(1.174)  &(1.610) &(2.206) &(3.717) &(5.188) &(7.702)\\
  &[1.611]&[4.079]&[5.313]&[6.188]&[6.415]&[6.544]\\
${E_{1}}$&3.966&5.931&8.420&14.52&20.39&30.37\\
  &(3.901)&(5.749)&(8.115)&(13.95)&(19.56)&(29.12)\\
  &[1.681]&[3.165]&[3.762]&[4.148]&[4.244]&[4.298]\\
${E_{2}}$&7.420&11.61&16.74&29.16&41.03&61.18\\
  &(7.382)&(11.54)&(16.64)&(28.98)&(40.78)&(60.81)\\
  &[0.523]&[0.612]&[0.6179]&[0.6157]&[0.6145]&[0.6138]\\
${E_{4}}$&16.15&26.48&38.73&68.01&95.90&143.2\\
  &(16.30)&(26.83)&(39.29)&(69.05)&(97.38)&(145.4)\\
 &[0.9170]&[1.302]&[1.426]&[1.499]&[1.517]&[1.527]\\
${E_{6}}$&26.88&45.08&66.36&117.0&165.1&246.5\\
   &(27.29)&(45.94)&(67.70)&(119.4)&(168.5)&(251.7)\\
   &[1.50]&[1.870]&[1.98]&[2.043]&[2.058]&[2.067]\\
${E_{10}}$&53.24&91.17&135.0&238.7&337.1&503.8\\
   &(54.31)&(93.26)&(138.2)&(244.5)&(345.3)&(516.1)\\
   &[1.967]&[2.245]&[2.323]&[2.367]&[2.377]&[2.383]\\ 
${E_{14}}$&85.01&147.0&218.3&386.6&546.2&816.3\\
  &(86.78)&(150.4)&(223.4)&(395.7)&(559.1)&(835.6)\\
  &[2.047]&[2.230]&[2.279]&[2.306]&[2.313]&[2.316]\\ 
${E_{17}}$&111.9&194.4&289.0&512.1&723.7&1082.0\\
  &(114.0)&(198.3)&(294.9)&(522.7)&(738.6)&(1104.0)\\
  &[1.868]&[1.974]&[2.001]&[2.016]&[2.020]&[2.022]\\    
\end{tabular}
\end{center}
\end{table}
\begin{table}
 {\bf Table- 4: }
Sample results  for the octic- AHO in the zeroth - order (GHA)  compared with results of earlier calculations from ref.[ 7 ] (shown in parentheses), over a wide range of $'\lambda'$
and $'n'$.
\begin{center}
\begin{tabular}{c c c c c c }
\multicolumn{1}{ c }{$\lambda= $}&
\multicolumn{1}{ c }{\quad$0.1$ }&
\multicolumn{1}{ c }{\quad$1.0$}&
\multicolumn{1}{ c }{\quad$5.0$}&
\multicolumn{1}{ c }{\quad$50.0$}&
\multicolumn{1}{ c }{\quad$200.0$}\\
${E_{0}}$   &1.3005&1.7794&2.3290&3.5565&4.6425\\
        &(1.2410)&(1.6413)&(2.1145)&(3.1886)&(4.1461)\\
${E_{1}}$   &4.4717&6.3946&8.5167&13.172&17.259\\
        &(4.2754)&(5.9996)&(7.9296)&(12.1950)&(15.9519)\\
${E_{2}}$   &8.6264&12.717&17.126&26.698&35.062\\
        &(8.4530)&(12.421)&(16.711)&(26.033)&(34.183)\\
${E_{4}}$   &19.763&30.026&40.863&64.165&84.444\\
        &(19.9930)&(30.4605)&(41.4947)&(65.20180)&(85.8251)\\
${E_{6}}$   &34.217&52.669&72.044&113.48&149.47\\
        &(35.0560)&(54.1403)&(74.0830)&(116.7629)&(153.8278)\\
${E_{8}}$   &51.570&80.013&109.65&172.99&227.97\\
        &(53.145590)&(82.6496)&(113.3486)&(178.9215)&(235.8193)\\
${E_{9}}$   &61.239&95.255&130.64&206.23&271.81\\
        &(63.2253)&(98.5529)&(135.2598)&(213.6157)&(281.5864)\\
${E_{10}}$  &71.532&111.49&153.01&241.64&318.52\\
        &(73.9545)&(115.4899)&(158.5991)&(250.5751)&(330.3433)\\
${E_{11}}$ &824.24&128.68&176.69&279.14&368.00\\
       &(85.3079)&(133.4201)&(183.3103)&(289.7106)&(381.9720)\\
${E_{12}}$ &93.893&146.79&201.65&318.67&420.14\\
       &(97.2636)&(152.3080)&(209.3443)&(330.9440)&(436.3695)\\
${E_{13}}$ &105.92&165.79&227.84&360.14&474.85\\
       &(109.7967)&(172.1125)&(236.6436)&(374.1834)&(493.4143)\\
${E_{14}}$&118.49&185.65&255.21&403.50&532.06\\
        &(122.8909)&(192.8082)&(265.1732)&(419.3737)&(553.0335)\\
\end{tabular}
\end{center}
\end{table}
\end{document}